\documentclass[conference]{IEEEtran}

\ifCLASSINFOpdf
   \usepackage[pdftex]{graphicx}
 
\else

   \usepackage[dvips]{graphicx}
 
\fi

\usepackage{algorithmic}

\usepackage{multirow}
\usepackage{amsmath, amssymb}
\usepackage{array}
\usepackage{subfigure}
\usepackage{amsmath}
\usepackage{pseudocode}

\begin{document}
\title{Diffusion-Aware Sampling and Estimation in Information Diffusion Networks}

\author{\IEEEauthorblockN{Motahareh Eslami Mehdiabadi}
\IEEEauthorblockA{Sharif University of Technology\\
Email: eslami@ce.sharif.edu}
\and
\IEEEauthorblockN{Hamid R. Rabiee}
\IEEEauthorblockA{Sharif University of Technology\\
Email: rabiee@sharif.edu}
\and
\IEEEauthorblockN{Mostafa Salehi}
\IEEEauthorblockA{Sharif University of Technology\\
Email:mostafa$\_$salehi@ce.sharif.edu}}

\maketitle

\begin{abstract}

Partially-observed data collected by sampling methods is often being studied to obtain the characteristics of information diffusion networks. However, these methods usually do not consider the behavior of diffusion process. In this paper, we propose a novel two-step (sampling/estimation) measurement framework by utilizing the diffusion process characteristics.
To this end, we propose a link-tracing based sampling design which uses the infection times as local information without any knowledge about the latent structure of diffusion network. To correct the bias of sampled data, we introduce three estimators for different categories; link-based, node-based, and cascade-based. To the best of our knowledge, this is the first attempt to introduce a complete measurement framework  for diffusion networks. We also show that the estimator plays an important role in correcting the bias of sampling from diffusion networks. Our comprehensive empirical analysis over large synthetic and real datasets demonstrates that in average, the proposed framework outperforms the common BFS and RW sampling methods in terms of link-based characteristics by about $37\%$ and $35\%$, respectively.

\end{abstract}

\IEEEpeerreviewmaketitle
\section{Introduction}

Information diffusion is one of the important topics that has been considered in large On-line Social
Networks (OSN) such as Facebook, Twitter, and YouTube. These networks that provide information in different formats such as posts, tweets, and videos are called ``information diffusion networks". 
In recent years, the tremendous growth of these networks, have resulted in creation of large information networks. For example, in March 2011, Twitter users were sending 50 million tweets per day \cite{twitter}.
Moreover, the latent structure of diffusion networks makes their analysis considerably difficult. Although we usually discover the time of obtaining some information by people, we can not find the source of information easily. Furthermore, in epidemic diseases, the infection shows itself when somebody becomes infected without determining who infected whom \cite{Gomez10}. Therefore, it may be impossible or costly to obtain the complete structure of a large and latent diffusion network.

Partially-observed network data is often being studied to obtain the characteristics of these
networks. The network resulting from such measurements may be thought of as a sample from a larger underlying network. As a result, the accuracy of the studies on diffusion network analysis depends on the estimation of the characteristics based on the sampled network data.

The measurement of the network characteristics can be achieved in two steps: 1) Sampling, and 2) Estimation. In the first step, data is collected from the network by using a sampling method. The essential property of a sampling method that makes it appropriate for network inference is that its visiting probabilities should be known for all the network elements. This allows sampled data to be weighted so that they accurately represent the network data. In the estimation step, an estimator is used to obtain the network characteristics. An estimator is a function that uses a summary of sampled data as input, and estimates the unknown parameters of the population which has generated the input. However, sampling and estimation in the context of networks may introduce some potential complications.

\begin{figure*}[htp]
  \begin{center}
    \subfigure[Common Network Sampling Design]{\includegraphics[scale=0.16]{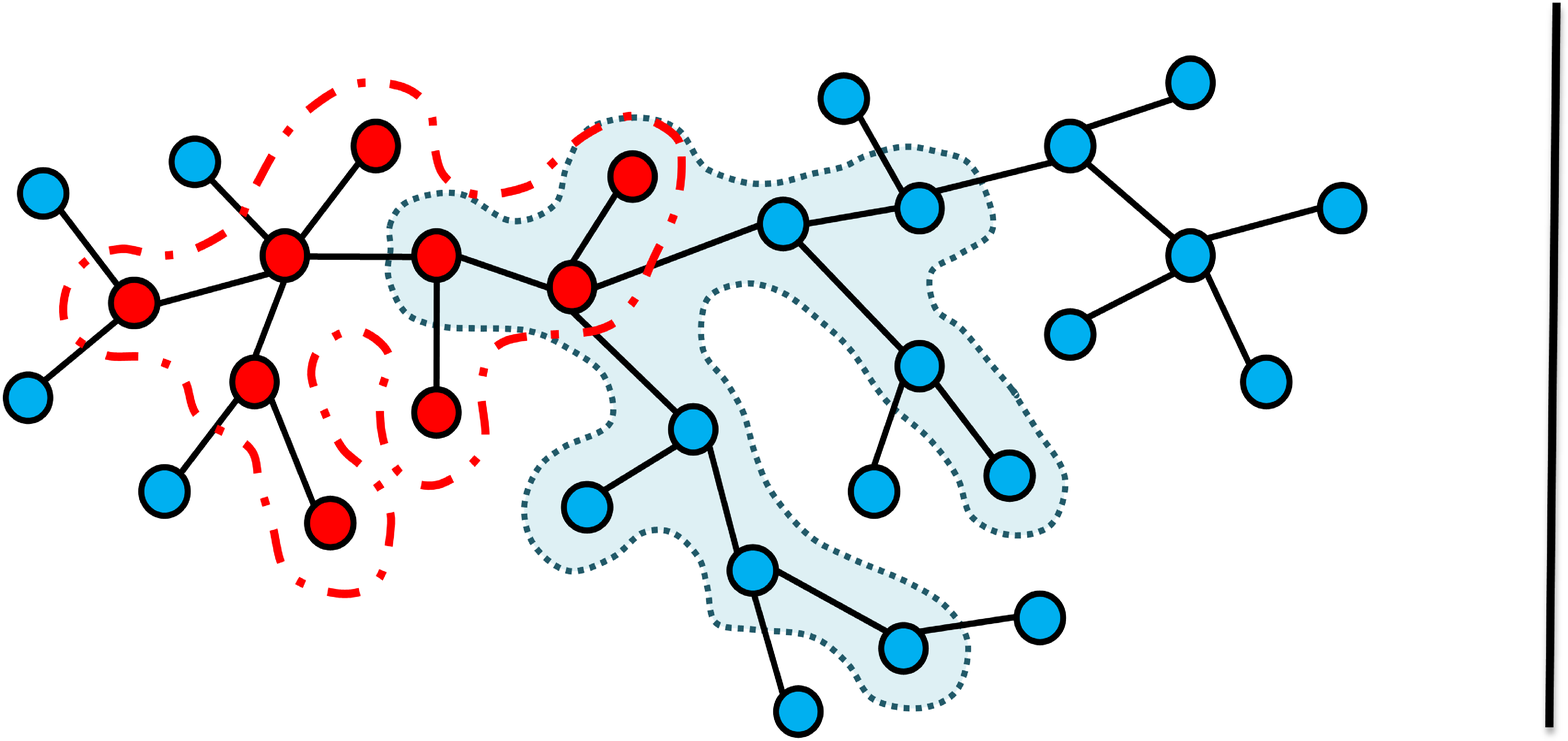}}
    \subfigure[Diffusion-Aware Network Sampling Design] {\includegraphics[scale=0.09]{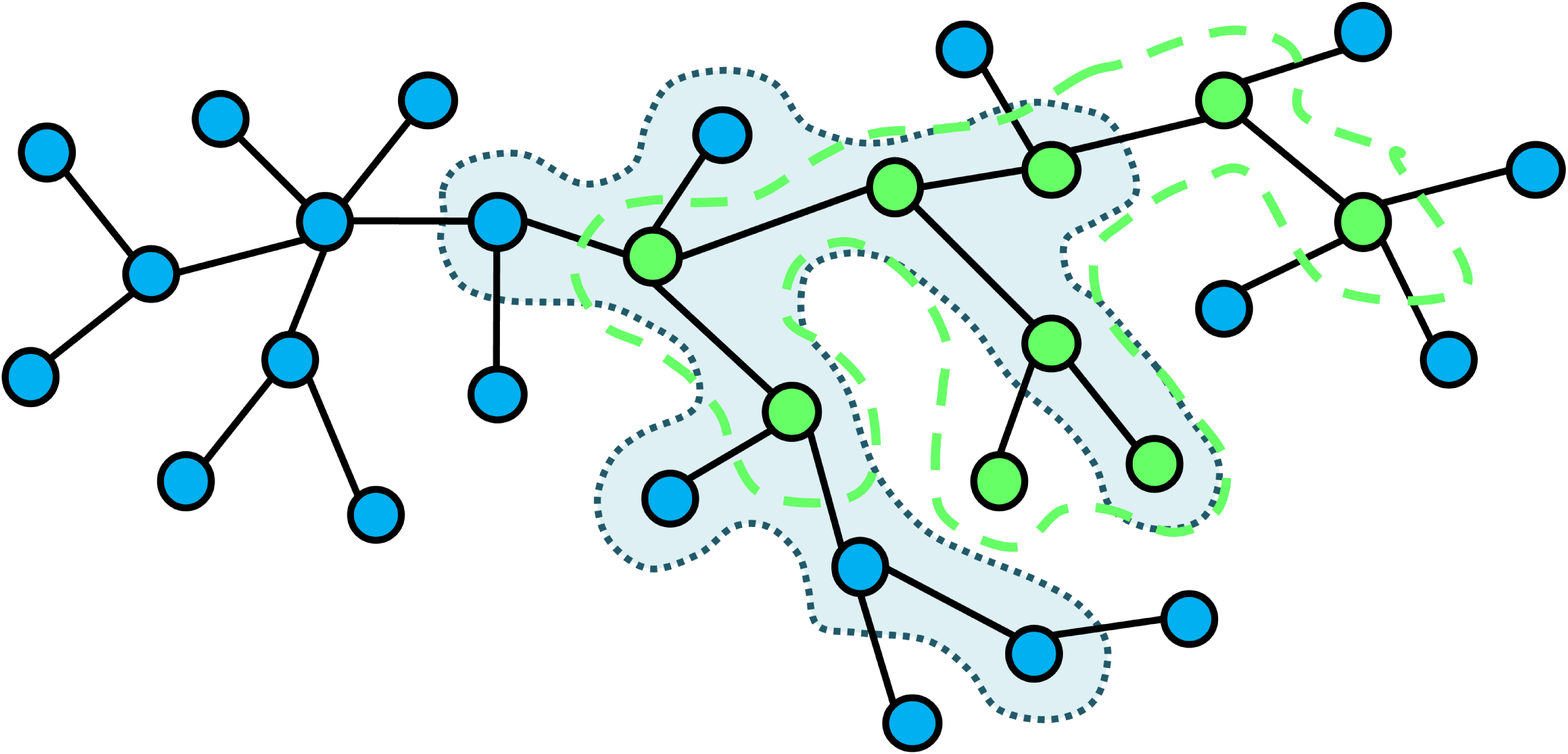}}
  \end{center}
  \caption{Illustration of different sampling designs in diffusion networks. The regions specified by dotted lines show the diffusion networks. The red and green areas demonstrate the sampled networks obtained by common and diffusion-aware network sampling methods, respectively. As it is shown, diffusion-aware network sampling design can cover the diffusion network more accurately.}
  \label{fig:Comparison}
\end{figure*}

In recent years, a considerable amount of research have been done on analyzing the topological characteristics of large OSNs based on the sampled data from different networks such as Facebook \cite{gjoka2010, gjoka2011}, Twitter \cite{Salehi2011}, YouTube \cite{mislove2007}, and other large networks \cite{Leskovec2006, Salehi12}. However, considering the sampling approaches to study diffusion behaviors of social networks, apart from their topologies, is a remarkable issue that should be addressed. 
The previous work on diffusion data collection \cite{Choudhury10, Eslami12, Sadikov11, Lin11} have used some well-known sampling methods such as Breadth-First Search (BFS) and Random Walk (RW), without considering the behavior of the diffusion process. 
This leads to gathering redundant data and losing parts of diffusion data, that consequently decrease the performance of these sampling methods (refer to Figure \ref{fig:Comparison}). On the other hand, often it is not feasible to directly work with the diffusion networks, because the structure of many large real systems can not be discovered. 
Moreover, the previous studies assert that the characteristics of a sampled diffusion network is indicative of the same characteristics for the whole network. However, it should be noted that the obtained characteristics represent the sampled graph, instead of the original graph. Such problems can be compensated for in many cases by using the appropriate estimator.

In this paper, we propose a novel two-step (sampling/estimation) framework, called ``\textsc{Dns}", to measure the characteristics of diffusion networks. To this end, we propose a link-tracing based sampling method that utilizes diffusion process properties to traverse the network more accurately. Specifically, this method samples the underlying network by moving from a node to one of its neighbors through an outgoing link based on the probability of spreading infection. We calculate this infection probability by considering the cascades behavior in the diffusion networks. It is noteworthy that the algorithm only uses the infection times as local information without any knowledge about the latent structure of the diffusion network. Moreover, we extend the well-known Hansen-Hurwitz estimator \cite{Hansen1943} to correct the bias of sampled data. We propose three efficient estimators related to different categories of network characteristics; link-based, node-based, and cascade-based. To the best of our knowledge, this is the first attempt to introduce a complete measurement framework for the diffusion networks.

We evaluated the proposed framework over large synthetic and real datasets by comparing it with BFS and RW sampling methods. The experimental results demonstrated that \textsc{Dns} outperforms the aforementioned common sampling methods in terms of link-based characteristics by about $36\%$, in average. Moreover, \textsc{Dns} decreased the bias of the sampled data by $30\%$ compared to the sampling design without estimation. The results confirm that finding an appropriate estimator has an important role in correcting the bias of sampling methods. Furthermore, the results show that the proposed framework performs well even in low sampling rates. 
we also analyzed the effect of diffusion rate on the performance of \textsc{Dns}. The analysis showed the independence of the proposed framework to the diffusion process behavior. Hence, we can use \textsc{Dns} in various diffusion networks with different diffusion patterns without any performance loss.

In summary, our main contributions can be summarized as follows:
\begin{itemize}
\item
Proposing a novel sampling design for gathering data from a diffusion network by utilizing the properties of diffusion process.
\item
Proposing three estimators for correcting the bias of sampled data by computing the visiting probabilities of different types of diffusion characteristics (link-based, node-based, and cascade-based).
\item
Decreasing the bias of measuring link-based characteristics compared to the other common sampling methods

\end{itemize}

The rest of the paper is organized as follows. Section \ref{Related Work} presents a classification of data collection approaches in the field of information diffusion networks. The problem formulation is proposed in Section \ref{sec:prob formulation}. The proposed measurement framework is presented in Section \ref{sec:Proposed Framework}. Section \ref{Experimental Evaluation} elaborates the experimental evaluation, and the concluding remarks are provided in Section \ref{Conclusions}.

\section{Related Work} \label{Related Work}
Diffusion process as a fundamental phenomenon over OSNs has attracted great attention in recent years \cite{Gomez10, Myers10, gruhl04, Kossinets08, Nowell08, Leskovec09, Yang10, Sadikov11, Eslami11}. Here, we provide a comprehensive survey over the approaches used for collecting the diffusion process data.

\textbf{Complete Data:} The most fundamental approach is to collect the complete diffusion data. Many diffusion processes try to generate some diffusion paths and use them for analysis. Following the Iraq war petitions in the format of e-mail \cite{Nowell08, Chierichetti11}, studying communication events between faculty and staff of a university by e-mails \cite{Kossinets08}, and tracking the flow of information by extracting short textual phrases \cite{Leskovec09}  are some examples of this approach. However, gathering diffusion data in many areas create problems such as missing data, privacy policies, and impossibility of tracing all paths of diffusion. Moreover, large scale of diffusion networks is one of the most important obstacles of gathering the complete diffusion data. These problems have led the researchers to use sampling methods to obtain partial diffusion data. 

\textbf{Partial Data:}
Sampling methods can be considered as an efficient way to tackle the problem of large-scale diffusion data. Using these methods to collect diffusion data have been studied in some recent work \cite{Choudhury10, Eslami12, Sadikov11, Lin11}. The majority of these works have utilized one of the most common sampling methods; Breadth-First Search (BFS). BFS is a basic graph-based sampling method that has been used extensively for sampling networks in various domains \cite{mislove2007, gjoka2010, Wilson09, Salehi2011}. At each iteration of BFS, the earliest explored node is selected next. This method discovers all nodes within some distance from the starting node. Inferring diffusion topics from the DBLP database \cite{Lin11} and sampling the Twitter network to study on the resulting diffusion network \cite{Choudhury10, Sadikov11} are some examples which use BFS to collect the diffusion data.
However, BFS leads to a bias towards high degree nodes \cite{Becchetti2006}, and this bias has not been analyzed for arbitrary graphs \cite{kurant2010}. Despite the popularity of BFS, the problem of computing the visiting probabilities of network elements (such as nodes and links) in BFS sampling design is still unsolved. Because, sampling without replacement in BFS introduces complex dependencies between the sampled elements. To the best of our knowledge, no estimator has been introduced to correct the sampling bias of BFS in an arbitrary network.

Despite the considerable amount of research on analyzing the topological characteristics of the networks in various areas \cite{gjoka2010, gjoka2011, Salehi2011, mislove2007, Leskovec2006, Salehi12}, little attention have been made on gathering partial data based on the diffusion behavior. Random Walk (RW) \cite{Lovas93} is also one of the most important and widely used link-tracing sampling methods in different kind of network contexts such as uniformly sampling Web pages from the Internet \cite{henzinger2000}, content density in peer-to-peer networks \cite{Gkantsidis2006, Stutzbach2008}, degree distributions of the Facebook social graph \cite{gjoka2010, gjoka2011} and in general large graphs \cite{Leskovec2006}. A classic RW samples a graph by moving from a node $u$, to a neighboring node $v$, through an outgoing link $(u,v)$, chosen uniformly at random from the neighbors of node $u$. The probability of selecting the next node determines the probability that nodes are being sampled. In any given connected and non-bipartite graph $G$, the probability of being at a node $u$ converges at equilibrium to the stationary distribution $\pi(u)= deg(u)/ 2|E|$, where $deg(u)$ and $E$ are the degree of node $u$ and are the set of links of the network graph. Moreover, the probability that a link is visited is $1/|E|$ (i.e., links are visited uniformly at random) \cite{Lovas93}.

Using these sampling methods without any attention to diffusion paths will result in some redundant data which are not related to the diffusion process. Removing these unnecessary data decreases the efficiency of these sampling methods \cite{Eslami12}. No work has previously been done that considers diffusion process characteristics in the sampling strategy. In this paper, we propose a diffusion-aware sampling and estimation methods which uses only local information of the underlying network.
To the best of our knowledge, this is the first study to introduce a complete measurement framework for the diffusion networks. Moreover, we use BFS and RW as baseline methods for comparison.

\section{Problem Formulation} \label{sec:prob formulation}

\subsection{Basic Notations and Definitions}
\label{sec:notation}
Let network $G = (V, E)$ be the underlying network where $V$ is the set of nodes, and $E$ is the set of links where $n=|V|$ and $m=|E|$. In diffusion process, some diffusible chunks such as information and epidemic diseases propagate over $G$. These diffusible chunks are called ``infection" where each path of infection will build a ``cascade" \cite{Gomez10, Eslami11}. When the cascades spread over the underlying network, the diffusion network $G^*$ will be formed. 

We define $G_{s} = (V_{s}, E_{s})$ as the induced sub-graph of $G$ by sampling rate of $\mu$ where $V_{s} \subset {V}$ and $E_{s} \subset {E}$. In order to analyze the diffusion process, we should measure the diffusion characterization metrics from the sampled diffusion data. Since diffusion phenomenon covers many elements of the network (such as nodes, links, and cascades), we determine an ``element set", $T$, as a set of diffusion network elements \cite{Eslami12}. Let $L$ be a finite set of element labels. A label can be, for instance, the degree of a node, the weight of a link, or the length of a cascade. A label $l_{e}$  is assigned to each element $e \in T$ by a target function $f:T \rightarrow L$, i.e. $f = \{(e, l_{e})| e \in T, l_{e} \in L \}$. For example, infection is a label for each node that shows whether this node is infected during the diffusion process or not. The target function $f$ for this label will match nodes $u \in V$ to the set  $L=\{0, 1\}$ ($f(u) = 0$, if node $u$ is not infected and $f(u) = 1$, otherwise).

Almost all network characterization metrics we are aware of can be expressed as some aggregative function.
In this paper, we focus on the measurement of diffusion network characteristics. To this end, we consider the average function ($\eta$) over diffusion elements as:
\begin{equation} \label{probDef.}
\eta (f(G))=\frac{\sum_{e\in T} {f(e)} }{|T|}
\end{equation}
In the above infection example, this average shows the percentage of infected nodes by the diffusion process to all the nodes of the underlying network.

\subsection{Problem Definition}

Our goal is to propose a diffusion-aware measurement framework to collect diffusion data in an efficient way. The diffusion process measurement procedure consists of two steps: (1) samples from the underlying network and computes the desired target function $f$ on the sampled elements , (2) computes an estimate of $Avg (f)$ by finding an appropriate estimator \textit{M}. To evaluate the measurement framework, we define the bias metric as:
\begin{equation}
\rho = \frac{{|\eta(f({G^{*}})) - \eta(f(M))| }} {\eta(f({G^{*}}))}
\end{equation}
Now, our problem becomes equal to finding a measurement framework which minimizes the bias, i.e. $\rho$.

\section{Proposed Framework} \label{sec:Proposed Framework}

In this section, for the first time, we propose a diffusion-aware probabilistic measurement framework, called ``\textsc{Dns}" (\textbf{D}iffusion \textbf{N}etwork \textbf{S}ampling).

\subsection{Sampling Design} \label{DNS_Sampling}

In the existing sampling methods such as BFS and RW, we begin at a starting node, and recursively visit (one or more) of its neighbors as next nodes, without considering the diffusion paths. 
Here, we try to utilize the diffusion process properties to find how to traverse the network more accurately. By computing the probability of spreading infections over the links of underlying network, we can direct the sampling design toward diffusion paths without any prior knowledge about the diffusion network structures. Therefore, we can cover a greater part of unknown diffusion network and decrease the redundant data such as nodes and links which do not attend the diffusion process. 

To calculate the probability of spreading infection over a link, we focus on the cascades behavior in the diffusion networks. Each cascade $c$ can be assigned to a time vector $t_{c} = {\{t_{1},t_{2},\cdots,t_{n}\}}$ which shows the infection times of nodes by $c$. If cascade $c$ does not infect a node, this node infection time will be considered as $\infty$ \cite{Eslami11}. 
The cascades with the same structure that propagate over the underlying network is shown by set $C$ with $N_c$ members. We define $C_T = {\{t_{1},t_{2},\cdots,t_{N_c}\}}$ as the set of cascades' time vectors. 
We assume the transmission model of cascades follows the independent cascade model \cite{Kempe03}. In this model, a node gets the chance to transmit information to its neighbors at each time episode, independently. 

When a node decides to infect one of its neighbors, it will do the transmission with a waiting time model that shows how long it will take for a node to infect a chosen neighbor. In the proposed sampling method, we use the exponential model \cite{Gomez10} as the waiting time model. By defining $\Delta = t_{v} - t_{u}$, the infection transmission probability over link $e(u, v)$ at cascade $c$ can be computed as follows.\begin{equation}
P_{c}(e) = e^{-\frac{\Delta}{\alpha}} \qquad \text{Exponential Model}
\end{equation}
Where $\alpha$ is an adjustment parameter which determines how fast a cascade spreads. As it can be seen, the probability of spreading an infection have an inverse relation with $\Delta$. It is the symptom of a simple fact; when you receive an interesting E-mail, the passing of time will decrease the probability of forwarding it to your friends. 
Since diffusion network contains many cascades, each link $e(u,v)$ can attend more than one cascade. Therefore, $C_{e}$ is defined as the set of cascades which pass over link $e$. Now, we define for each link $e$ the infection probability $P_{e}$, by calculating its average probability over the attended cascades as:\begin{equation}\label{edge_prob}
P_{e} = \frac{\sum_{c \in C}{P_{c}(e)}}{|C_{e}|}
\end{equation}
The pseudo code of the proposed sampling design is shown in Algorithm \ref{DNS-algorithm}. This method samples the underlying network by moving from a node $u$, to a neighboring node $v$, through an outgoing link with the infection probability $P_{e}$. 
It is noteworthy that the algorithm only uses the infection times (i.e. $C_T$) as local information without any prior knowledge about the latent structure of the diffusion network.
\begin{pseudocode}[ruled]{The Sampling Design}{Seed, C_T,k, \alpha} \label{DNS-algorithm}
 	\mbox{\textit{v := Seed}} \qquad \mbox{\%\textit{v} is the current node }
	\\
	 \WHILE {(|E_{s}| < k)}  \qquad \mbox{\%\textit{k} is the sampling size}

	 \DO
		\BEGIN	
			\FOREACH u \in \mbox{Neighbors (v)} \DO
				\BEGIN
					e := (v, u)
					\\
					V_{s} \GETS V_{s} \cup u
					\\
					E_{s} \GETS E_{s} \cup e
					\\
					
					\FOREACH c \in C_{e}  \DO
						\BEGIN
							\Delta = t_{u} - t_{v}
							\\
							P_{c} (e) = e^{-\frac{\Delta}{\alpha}}
							\\
							P_{e} = P_{e} + P_{c} (e)
						 \END
						 \\
						
						 P_{e} = \frac{P_{e}}{|C_{e}|}
						 \\
						 v \GETS u \mbox{ with probability of } P_{e}
				\END
		\END
	\\

	G_{s} := (V_{s}, E_{s})
\\
\RETURN{G_{s}}
\end{pseudocode}

\subsection{Estimation Approach} \label{sec:Estimation}

The selection bias of a sampling method can be corrected by re-weighting of the measured
values. This can be done using the Hansen-Hurwitz estimator \cite{Hansen1943}, i.e. elements are weighted inversely proportional to their visiting probability. 
For any target function $f : T \rightarrow L$ that defines a characteristic (refer to Section \ref{sec:notation}), the estimator of Equation \ref{estimation_equation} provides an asymptotic estimate of the population mean $\mu$ of $f$ \cite{Volz2008}:
\begin{equation} \label{estimation_equation}
{\hat{\eta}}={{\sum\limits_{i=0}^{k-1}{{f(X_i) \over \pi(X_i)}}} \over {{\sum\limits_{i=0}^{k-1}{{1 \over \pi(X_i)}}}}}
\end{equation}

Where $X_i$ and $\pi(X_i)$ are the visited element (that could be nodes, links or cascades), and its visiting probability on the $i^{th}$ draw of sampling method, respectively. 
Therefore, to use this estimator, we should compute the probability of visiting each element in the proposed sampling procedure.
In the following, we address this issue and extend the above estimator for three different categories of elements; link-based, node-based, and cascade-based.

\subsubsection{Link-based Characteristics}\label{link-based}
The links have a great role in spreading infection over the networks. Gaining some information without having any connection to others for propagation, will be not valuable in a network. Therefore, link-based characteristics are the most important ones in the diffusion process. ``Link Attendance", as an example of link-based characteristics, shows the amount of presence in diffusion process for a link. The links with high attendance are significant in some applications such as finding potential paths of infection propagation in the epidemic spreading \cite{Eslami12}. 

Since in the proposed sampling method we move over links with the probability $P_{e}$ (Equation \ref{edge_prob}), the visiting probability of link $e$ will be equal to this probability; i.e. $\pi(e) = P_{e}$. We can use these visiting probabilities in Equation \ref{estimation_equation} to estimate the real value of link-based characteristics. As mentioned before, we only use the local knowledge to compute the visiting probabilities of the links. 

\subsubsection{Node-based Characteristics}
The number of ``Seeds" (the beginners of an infection) \cite{Eslami12} and ``participation" (the fraction of users involved in the information diffusion) \cite{Choudhury10} are some examples of node-based characteristics.
Diffusion process can be modeled as a Markov random walk over the underlying network $G$ (the details can be found in our previous paper \cite{Eslami11}). Therefore, the visiting probability of node $u$ in the proposed sampling method can be defined as:
\begin{equation}
\pi(u) = \sum_{v \in N(u)}{ \pi(v) \pi{(e_{vu})}}
\end{equation}

Where $N(u)$ is the set of node $u$' neighbors. The infection of node $u$ at time $t_{u}$ depends on the infection of its neighbors at the $t_{v}$ where $t_{v} < t_{u}$. If we define $\pi\ = \{\pi(0), \pi(1), \cdots, \pi(n-1)\}$, calculating $\pi$ needs the global knowledge of a network as it is the stationary distribution of the mentioned Markov chain. Since we do not have the global view of the network in sampling procedure, finding the exact value of $\pi$ is not possible in real systems. Therefore, finding an approximation of $\pi$ can be considered as a research direction in the future.

\subsubsection{Cascade-based Characteristics}
The cascades as the building blocks of a diffusion networks can determine many characteristics of a diffusion process. For instance, the depth of a spreading phenomenon can be determined by the length of its cascades \cite{Eslami12, Choudhury10}.
Owing to the fact that each cascade $c$ has a series of links which it spreads over them, its visiting probability depends on visiting all of its links \cite{Gomez10}. Therefore, we can define $\pi(c)$ as:
\begin{equation}
\pi(c) = \prod _{e \in c} {P_{c}(e)}
\end{equation}

This formula can be calculated by having all the links of a cascade. Since the probability of visiting a cascade needs the global knowledge of a network, it should be approximated by using local information.

\section{Experimental Evaluation}\label{Experimental Evaluation}
\subsection{Setup}
As discussed in Section \ref{sec:Estimation},
computing the visiting probability of network elements should be done by using the local information. Since calculating nodes and cascades visiting probability need global knowledge about the underlying network structure, we evaluate the Link-Attendance as a link-based characteristic. We use BFS and RW as baseline methods for comparison with \textsc{Dns}.

To build the diffusion network, many homogeneous cascades are generated with the same structure over the underlying network. The speed of cascades' transmission is determined by $\alpha$. To control the distance through which a cascade can propagates, we use the parameter $\beta$ \cite{Gomez10}. The fraction of the underlying network $G$ which is covered by the diffusion network $G^{*}$ is defined as the diffusion rate $\delta$.

\subsection{Dataset}
We utilize seven synthetic and real networks with different structures. The properties and cascade generation settings of the datasets are provided in Table \ref{Parameters}. 
\subsubsection{Synthetic Dataset}
We use the following models to generate synthetic data:
\begin{itemize}
\item
Forest Fire model \cite{LeKlFa05} is generated by the parameter matrix $[5;0.12;0.1;1;0]$ where entries illustrate the number of starting nodes, forward burning probability, backward burning probability, decay probability and probability of orphan nodes, respectively.
\item
The Kronecker graph\cite{LeskovecF07} with three different Kronecker parameter matrices are generated as: the Random graph \cite{Erdos60} (by Kronecker parameter matrix of $[0.9, 0.1; 0.1, 0.9]$), the hierarchical network \cite{Clauset08} ($[0.5, 0.5; 0.5, 0.5]$), and the Core-Periphery network \cite{LeskovecLDM08} by ($[0.9, 0.5; 0.5, 0.3]$).
\end{itemize} 

\begin{table}[h]
\caption{The Network and Cascade Generation Parameters.\label{Parameters}}
\begin{center}
\begin{tabular}[c]{|l|c|c|c|c|c|} 
\hline
{ \textbf{Network} } &  { \textbf{\textit{n}}} & {\textbf{\textit{m}}} & {\textbf{ $\alpha$}}  & {\textbf{ $\beta$}}  & { \textbf{ $\delta$}}\\
\hline
Forest Fire & $10000$ & $14305$ & $0.7$ & $0.5$ & $0.5$ \\ 
\hline
Core-Periphery &  $8192$ & $15000$ & $0.7$ &  $0.1$ & $0.5$  \\
\hline
Hierarchical  &  $8192$ & $11707$  & $0.4$ & $0.5$ & $0.5$  \\ 
\hline
Random(ER) &  $8192$ & $15000$ & $0.4$ & $0.4$ & $0.5$  \\
\hline
PolBlog & $1490$ & $19090$  & $1.3$ & $0.5$ & $0.5$ \\
\hline
Football & $115$ & $615$  & $0.6$ & $0.5$ & $0.5$ \\
\hline
NetScience & $1589$ & $2742$  & $0.6$ & $0.5$ & $0.5$\\
\hline
\end{tabular}
\end{center} 
\end{table}

\subsubsection{Real Dataset}
We used three real-world networks for evaluation purposes. The first network is based on links and posts of blogs in the political blogosphere around the time of the 2004 presidential election in US \cite{Adamic05}. The other network is a network of American football games between Division IA colleges during regular season of Fall 2000 \cite{Grivan02}. The last is co-authorship network of network theory scientists \cite{Newman06} which we is referred to as NetScience. 

\subsection{Speed of Cascade} 

\begin{figure}[h]
\centering
\includegraphics[scale=0.1]{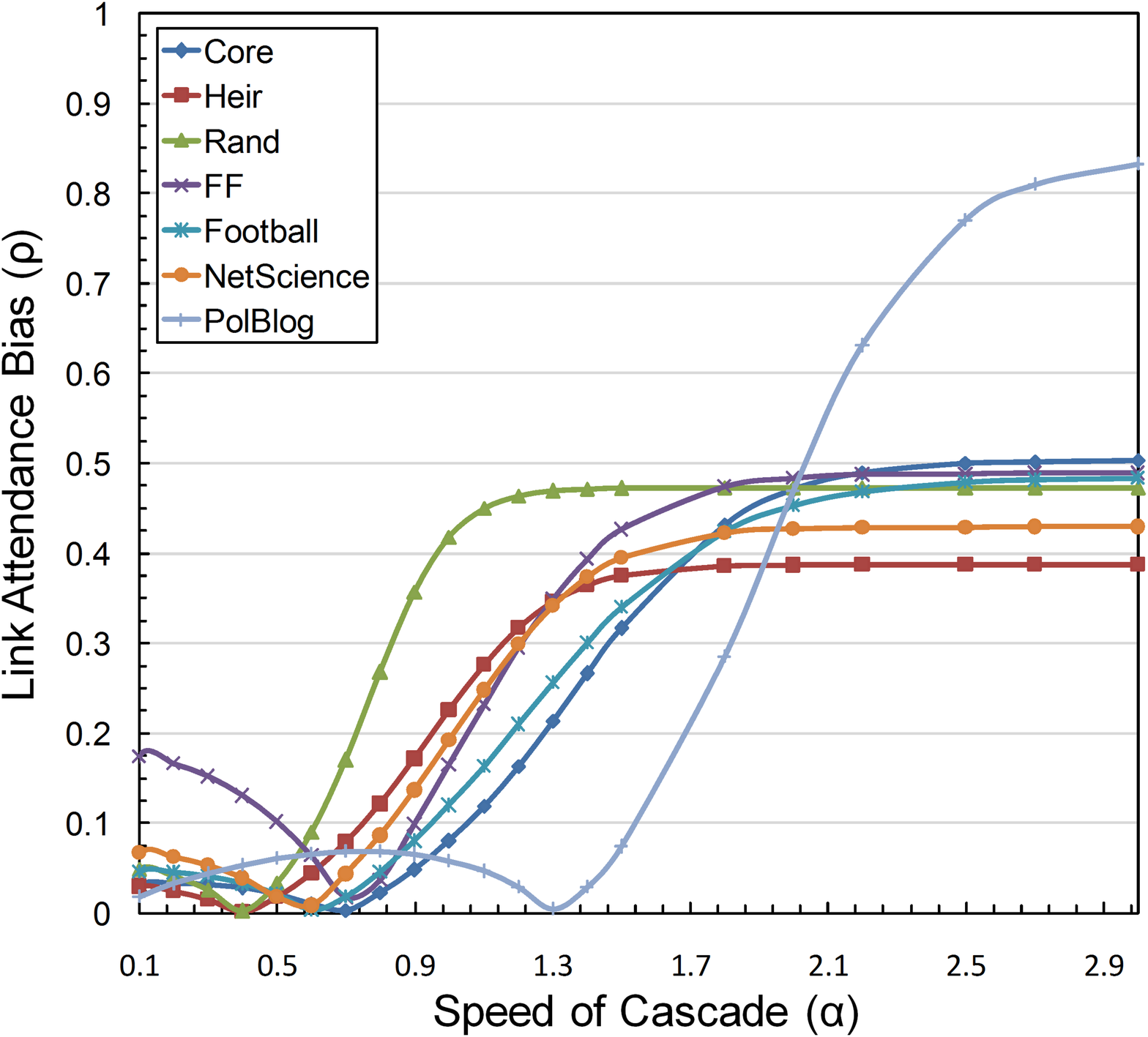}
\caption{\scriptsize{Speed of Cascade}}
\label{Alpha}
\end{figure}

As mentioned before, the speed of cascade propagation over the underlying network is controlled by $\alpha$. In the \textsc{Dns} framework, we use this parameter in calculating $P_{e}$ to determine the direction of sampling and correct the bias. As the diffusion network structure is unknown in the most large real systems, the speed of cascades is not available to be used in the sampling and estimation approach. Therefore, we evaluate the \textsc{Dns} performance by measuring the link-attendance characteristic based on different values of $\alpha$ ($0.1 < \alpha < 3$) over the synthetic and real networks in a fixed sampling rate ($\mu = 0.5$). As Figure \ref{Alpha} illustrates, all the networks have similar behavior with respect to $\alpha$. Moreover, most networks achieve the minimum bias (below $10\%$) in measuring the link-attendance characteristics when $0.4 \leq \alpha \leq 0.7$.

The behavior of the political blog network is different in comparison with other networks to some extent. Analyzing this network structure reveals that this different behavior is the result of the network density \cite{LeKlFa05} difference. 
Comparing the density of political blog network with the other networks shows that larger density needs larger $\alpha$ in the sampling and estimation procedure. In fact, the higher density necessitates more speed in cascade transmission to visit the elements of the network in a time episode. Therefore, political blog network achieves the least bias when $\alpha = 1.3$. 
The best value of $\alpha$ for each network is provided in Table \ref{Parameters}. We use these values as the input parameters for \textsc{Dns} in our experimental evaluations. 

\begin{figure*}[]
  \begin{center}
    \subfigure[Core Periphery Network]{\includegraphics[scale=0.055]{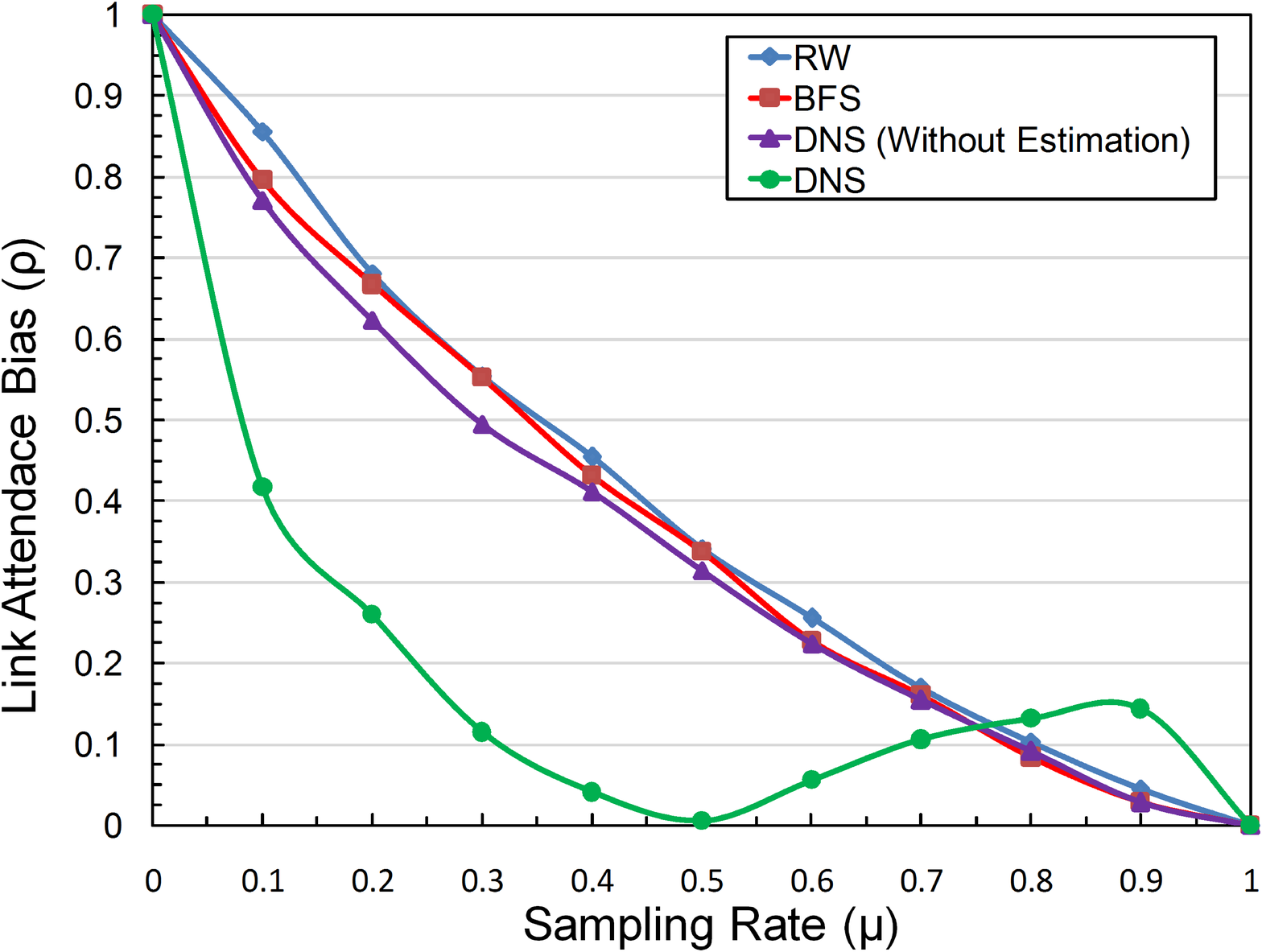}}
    \subfigure[Hierarchical Network] {\includegraphics[scale=0.055]{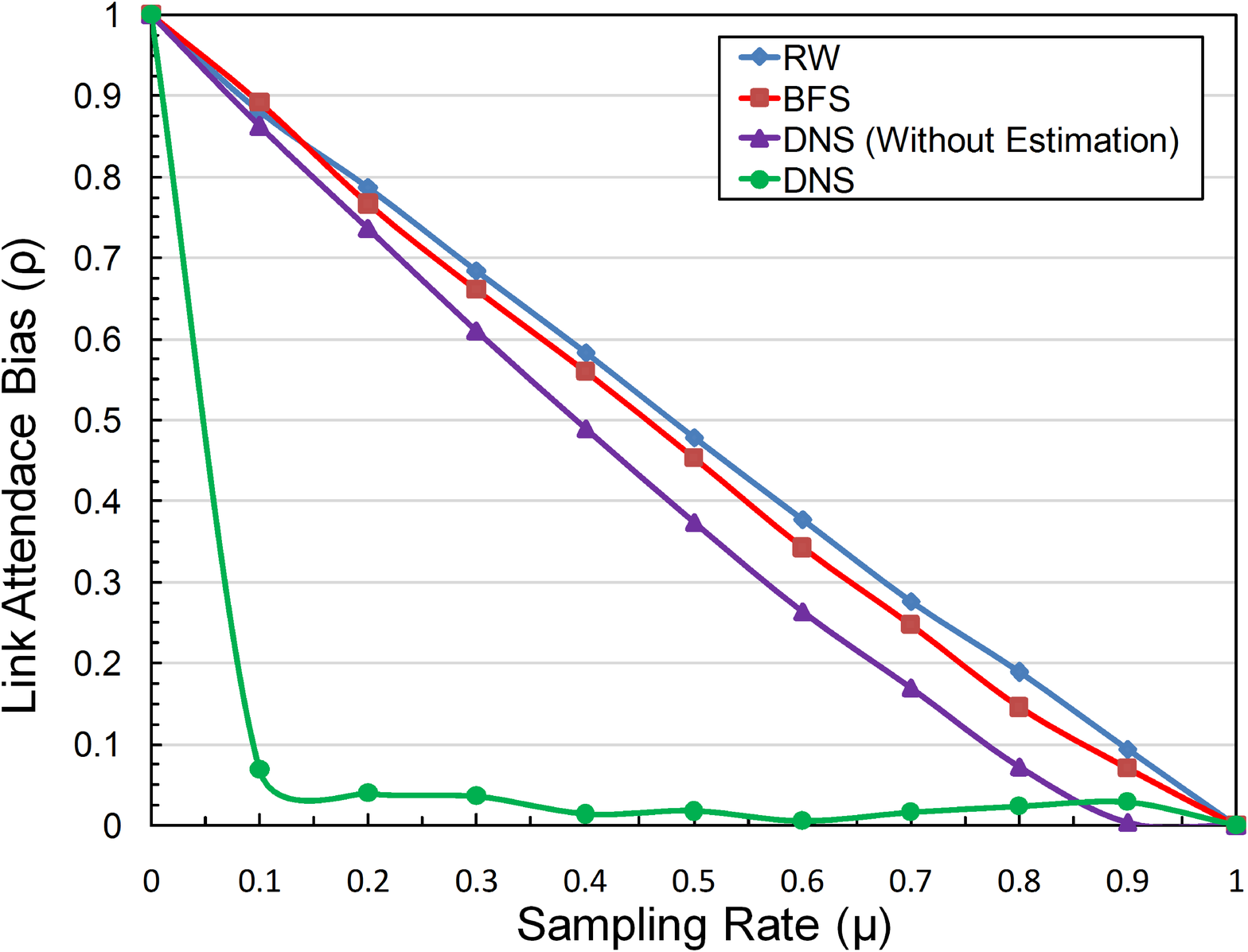}}
    \subfigure[Random Network] {\includegraphics[scale=0.055]{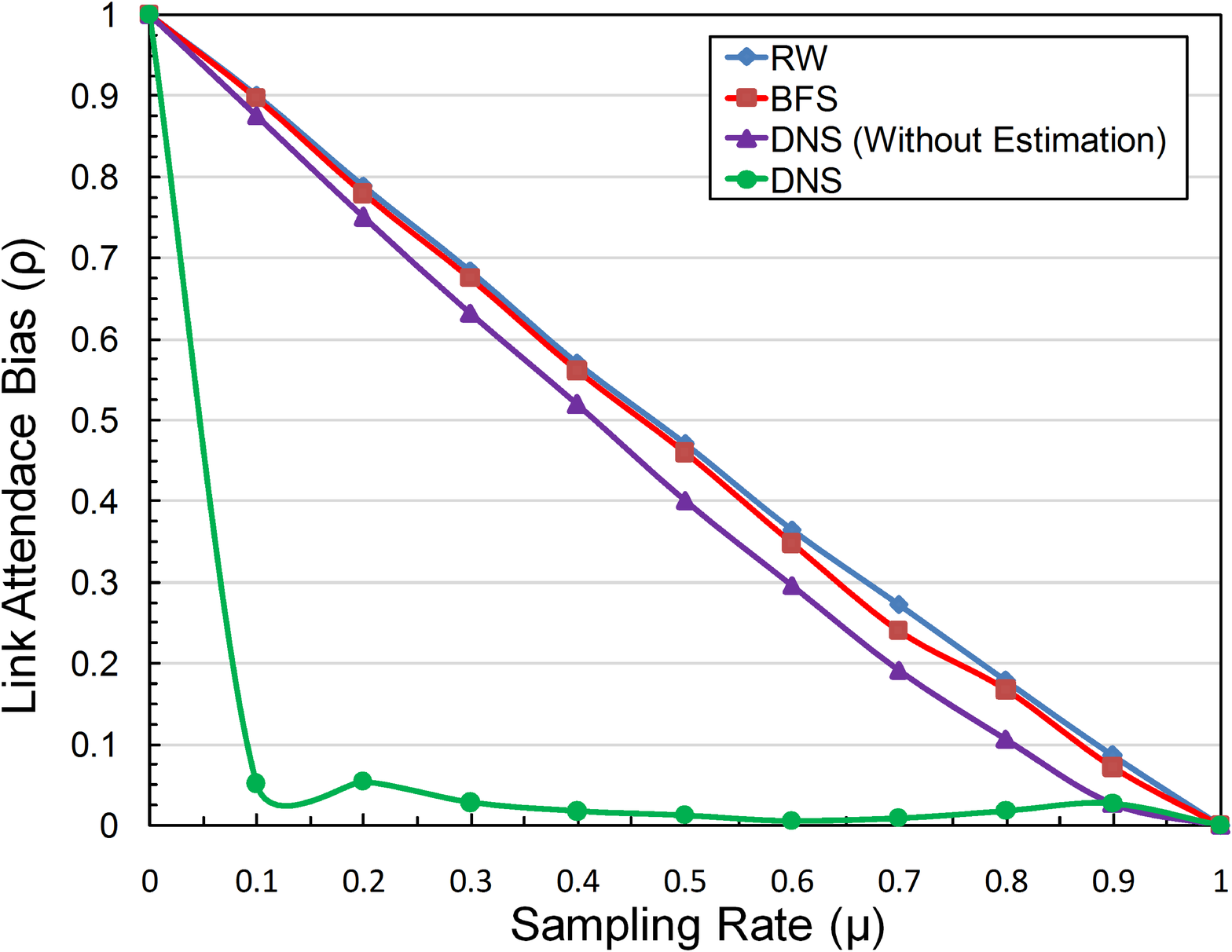}}
    \subfigure[Forest Fire Network]{\includegraphics[scale=0.055]{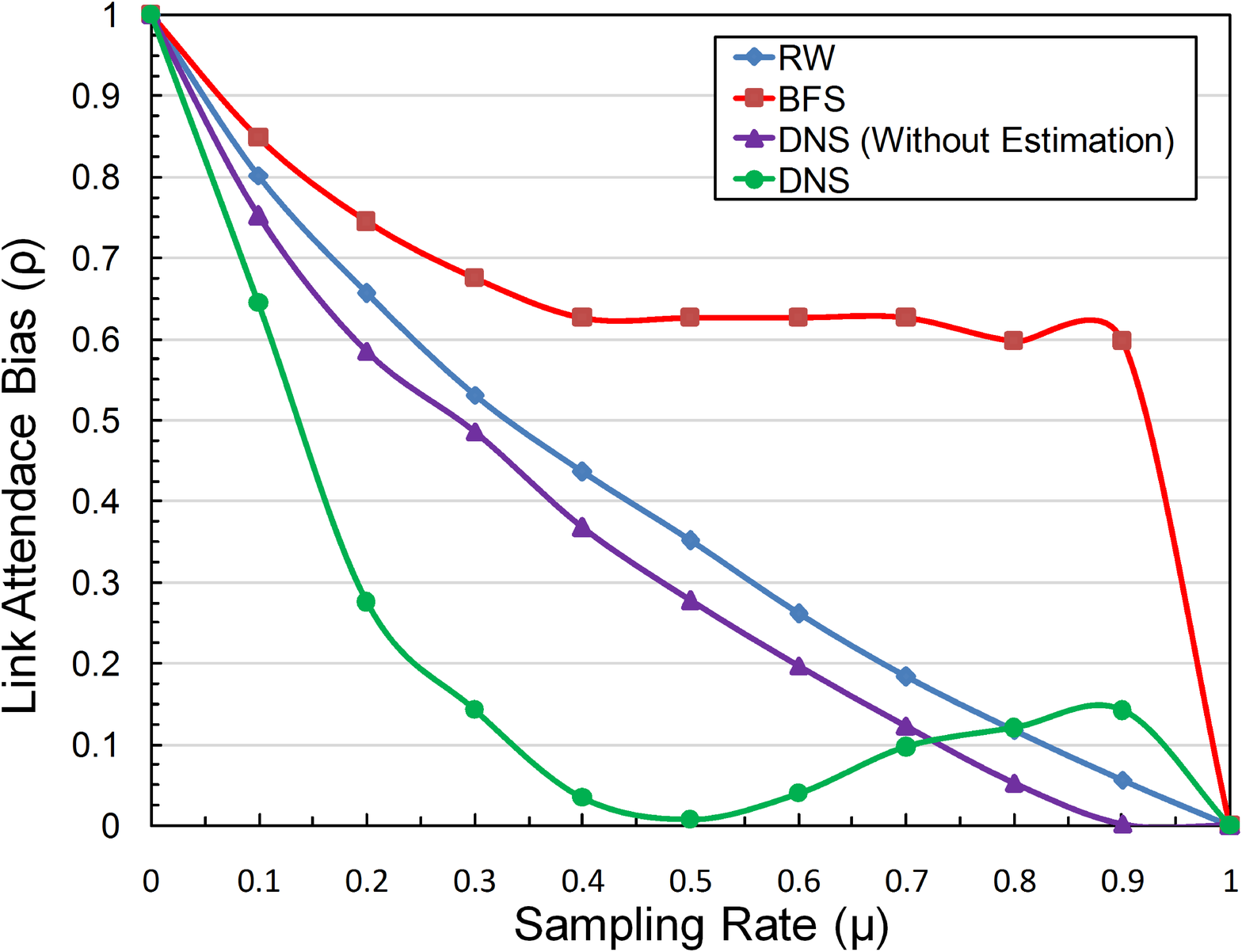}}
    \subfigure[Political Blogsphere Network] {\includegraphics[scale=0.06]{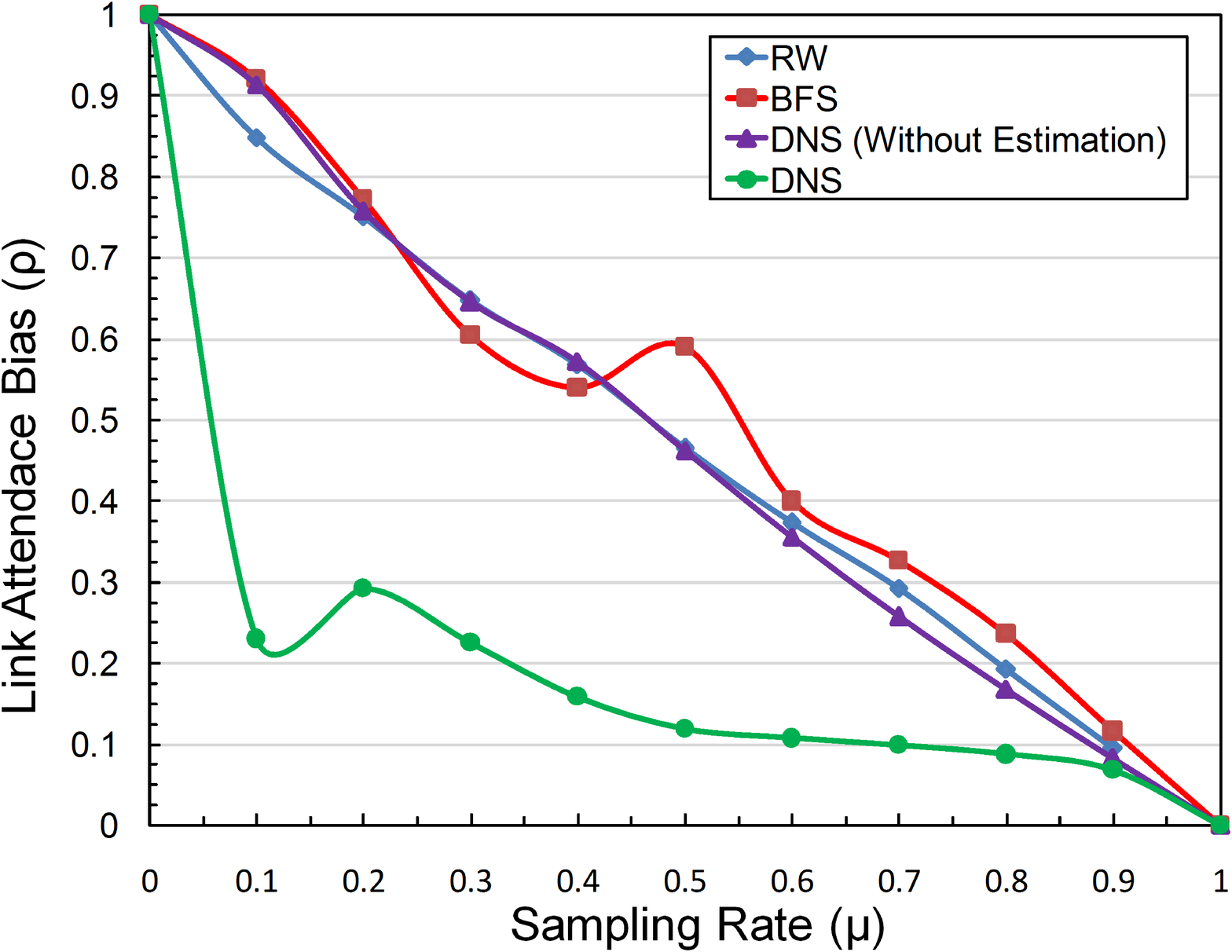}}
    \subfigure[Football Network] {\includegraphics[scale=0.06]{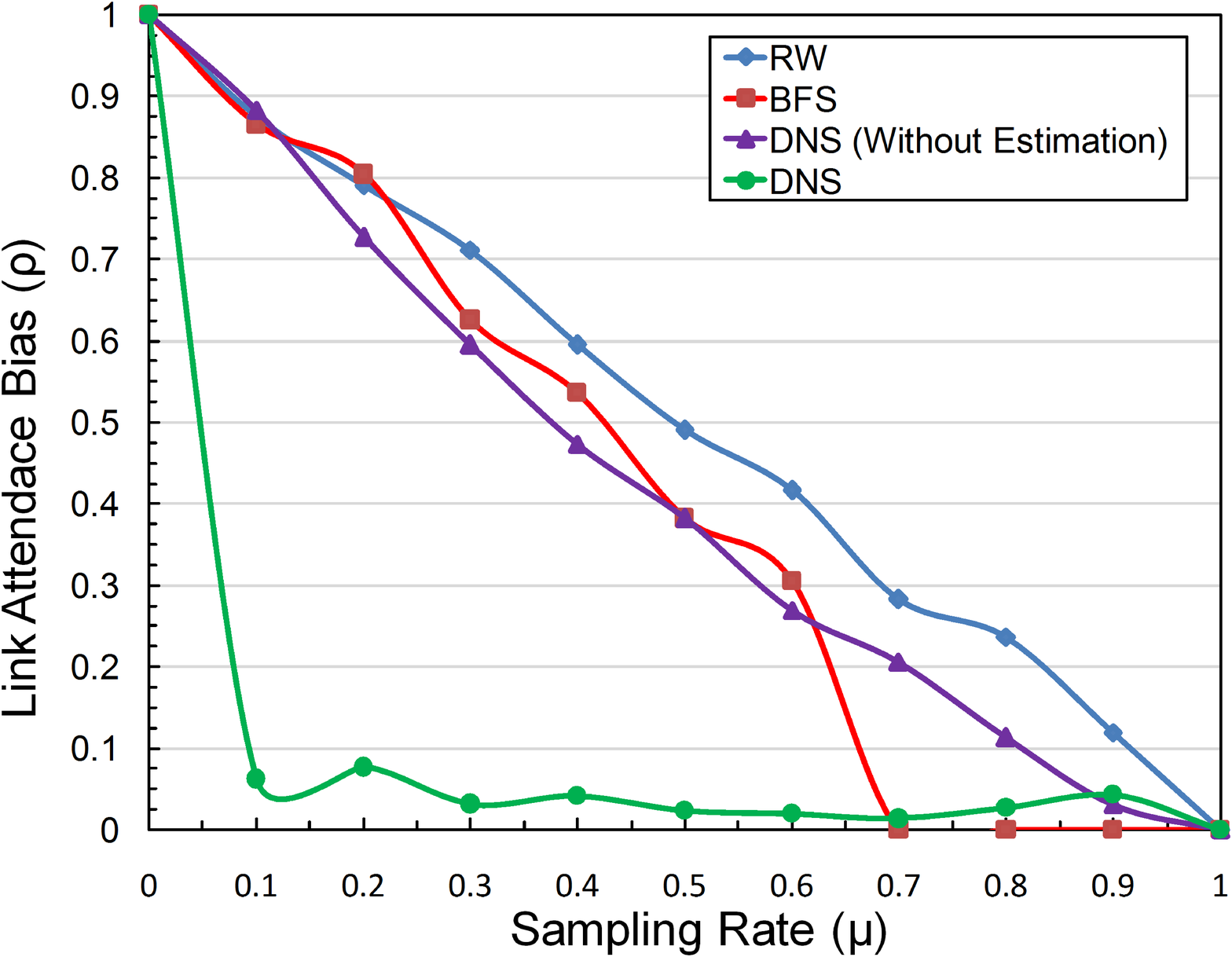}}
     \subfigure[Co-authorship Network]{\includegraphics[scale=0.06]{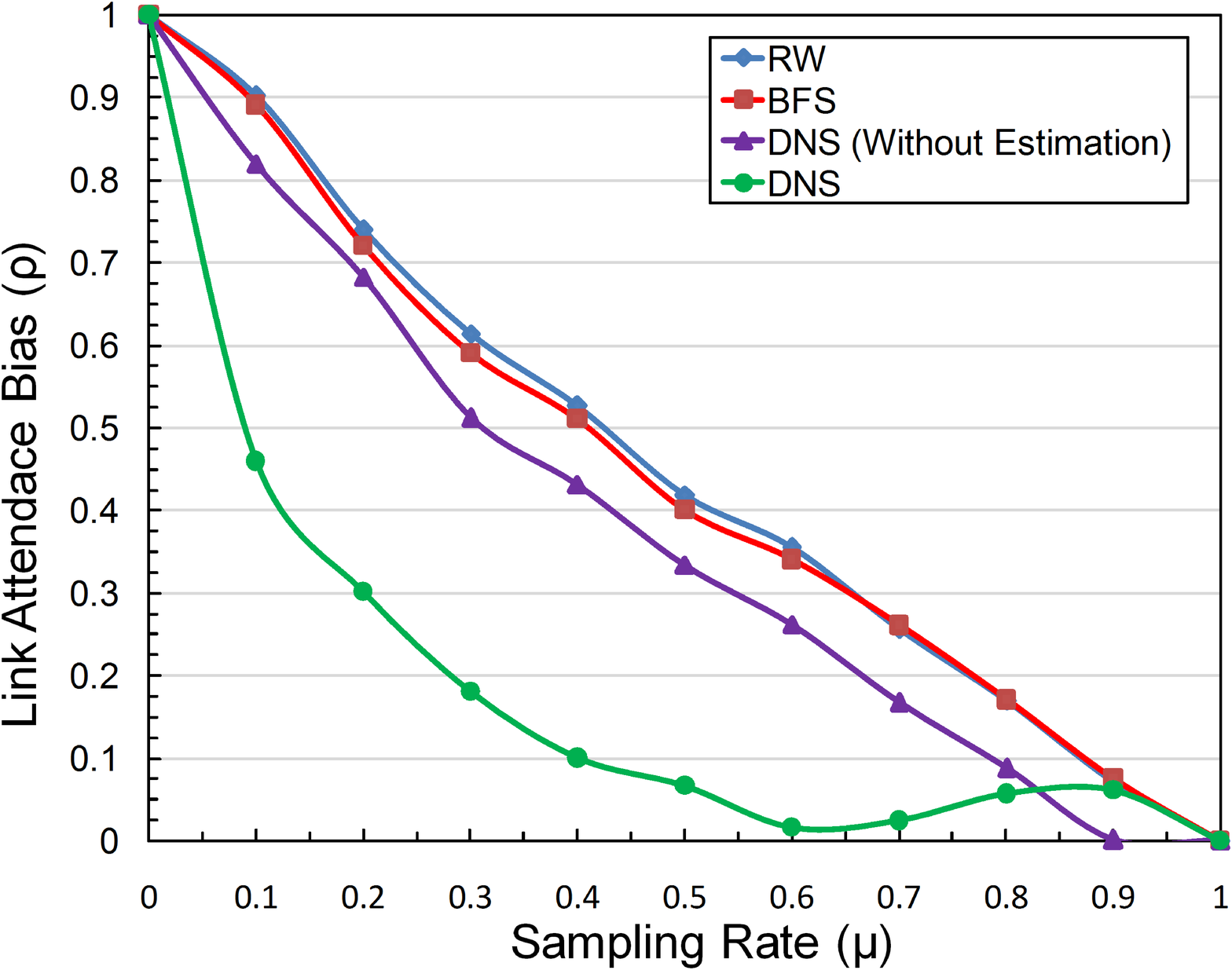}}
  \end{center}
  \caption{Link Attendance characteristic evaluation in different sampling rates}
  \label{Bias}
\end{figure*}

\subsection{Performance Evaluation}

In this section, we evaluate the performance of \textsc{Dns} framework in three aspects. First, we compare the bias of \textsc{Dns} with the baseline methods (BFS and RW) in measuring the link-based characteristics. Second, we study the importance of estimation approach in the proposed framework. Finally, we analyze the behavior of these methods in different sampling rates.

Figure \ref{Bias} shows the results of measuring the link-attendance bias against different sampling rates. As it is observed, the proposed framework can measure this characteristic with very low bias ($9\%$, in average). We summarize the average performance difference of \textsc{Dns} with BFS and RW in all networks in Table \ref{performance-difference}. It can be seen that \textsc{Dns} in average outperforms BFS and RW in terms of link-attendance by about $37\%$ and $35\%$, respectively.

Interestingly, we can see that the proposed framework has decreased the bias by $30\%$ compared to the sampling design of \textsc{Dns} without applying the proposed estimation approach. These results confirm that the obtained characteristics from a sampled data represent the sampled graph properties, but not the original graph. Therefore, an estimator plays an important role in correcting the bias of the sampling frameworks. However, this issue has not been considered in the previous work on gathering the diffusion data.
We also measured the node-based (Seed), and cascade-based (Depth) characteristics by the sampling design of \textsc{Dns} without estimation. The results show that the proposed sampling design alone, can not perform as good as \textsc{Dns} with estimation. Specifically, in average it can only improve the bias by about $12\%$, and $9\%$ compared to BFS and RW, respectively.

\begin{table}[h]
\small
\caption{\small{The average performance difference of \textsc{Dns} with BFS,   RW and \textsc{Dns} without estimation (\textsc{Dns}-WoE).}\label{performance-difference}} 
\begin{center}
\begin{tabular}[c]{|l|c|c|c|} 
\hline
{ \textbf{Network} }  & {\textbf{BFS}} & {\textbf{RW}} & {\textbf{\textsc{Dns}-WoE}} \\
\hline
Forest Fire & $49\%$ & $21\%$ & $14\%$  \\ 
\hline
Core-Periphery &  $22\%$ & $24\%$ & $20\%$ \\
\hline
Hierarchical  &  $43\%$ & $45\%$ & $37\%$ \\ 
\hline
Random(ER) &  $44\%$ & $45\%$ & $39\%$  \\
\hline
PolBlog & $34\%$ & $31\%$ & $31\%$  \\
\hline
Football & $35\%$ & $46\%$ & $37\%$  \\
\hline
NetScience & $30\%$ & $31\%$  & $22\%$ \\
\hline
\textbf{Average} & $37\%$ & $35\%$  & $30\%$ \\
\hline

\end{tabular}
\end{center} 
\end{table}

Moreover, Figure \ref{Bias} demonstrates that the proposed method can act very well even in low sampling rates. \textsc{Dns} decreases the bias of measuring diffusion characteristics to $3\%$ when $\mu < 0.3$. This promising result provides an appropriate sampling and estimation framework for the large real networks where only low sampling rates are available.

\subsection{Diffusion Behaviour Analysis}
\begin{figure*}[]
  \begin{center}
    \subfigure[Core Periphery Network]{\includegraphics[scale=0.055]{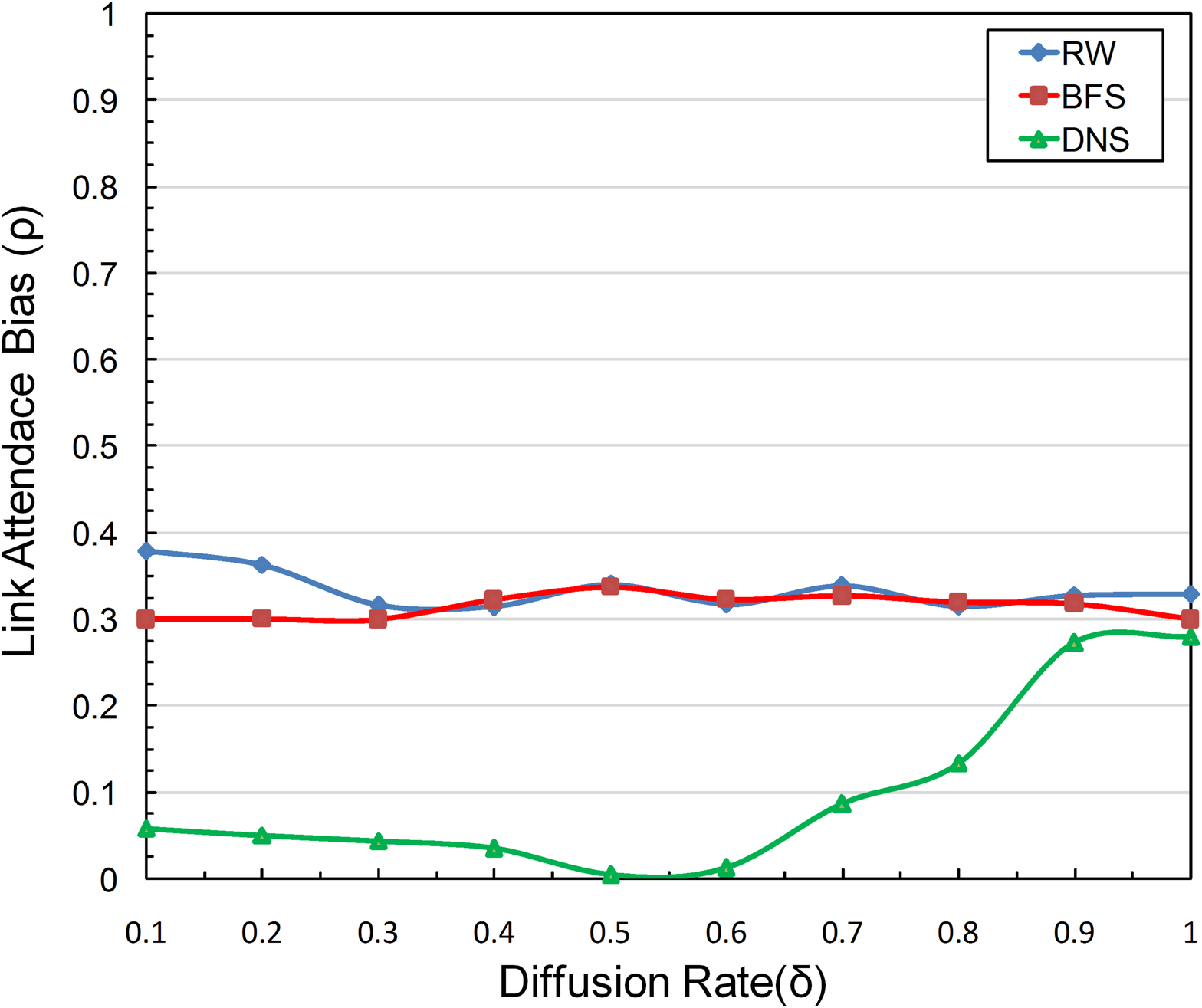}}
    \subfigure[Hierarchical Network] {\includegraphics[scale=0.055]{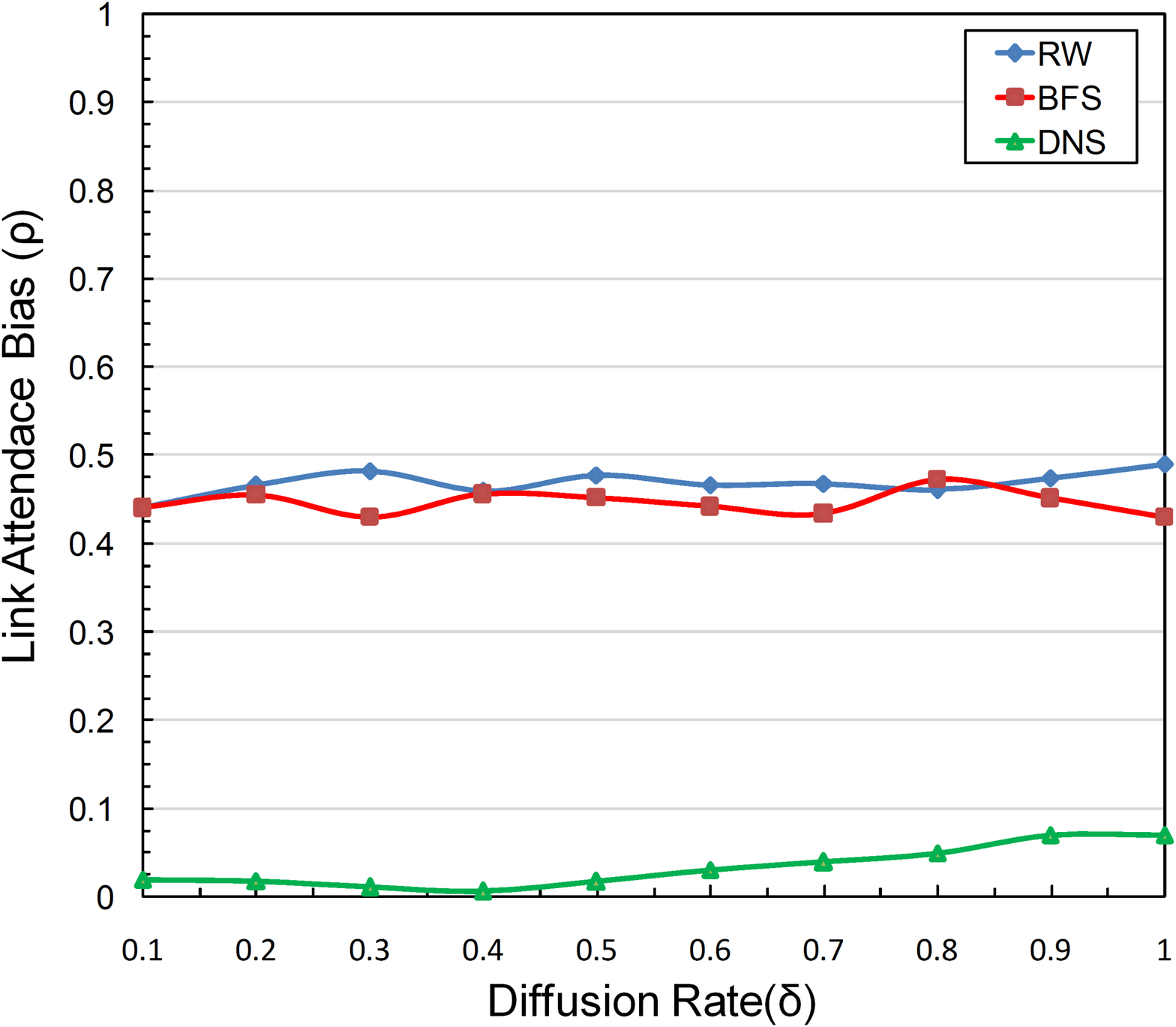}}
    \subfigure[Random Network] {\includegraphics[scale=0.055]{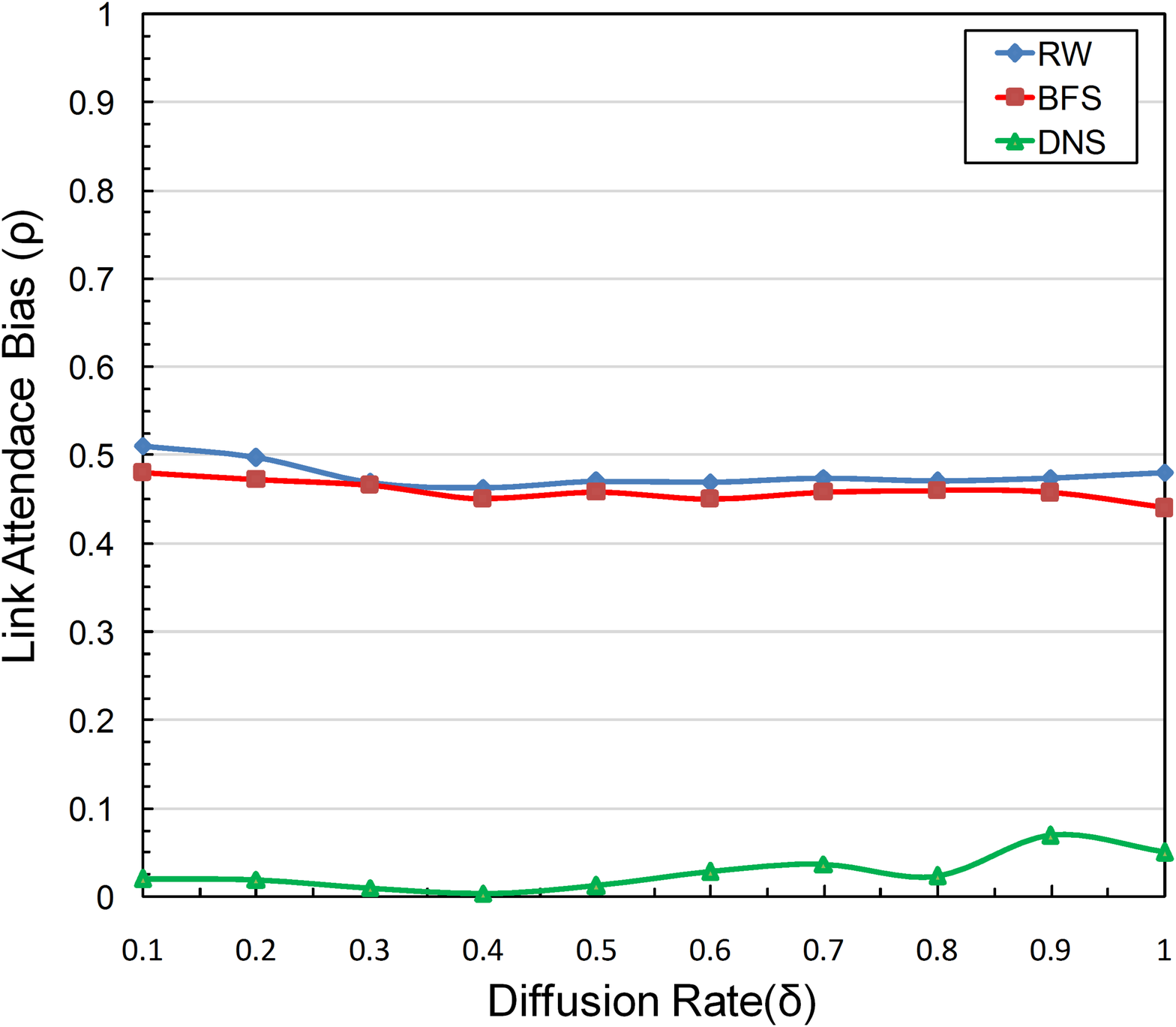}}
    \subfigure[Forest Fire Network]{\includegraphics[scale=0.055]{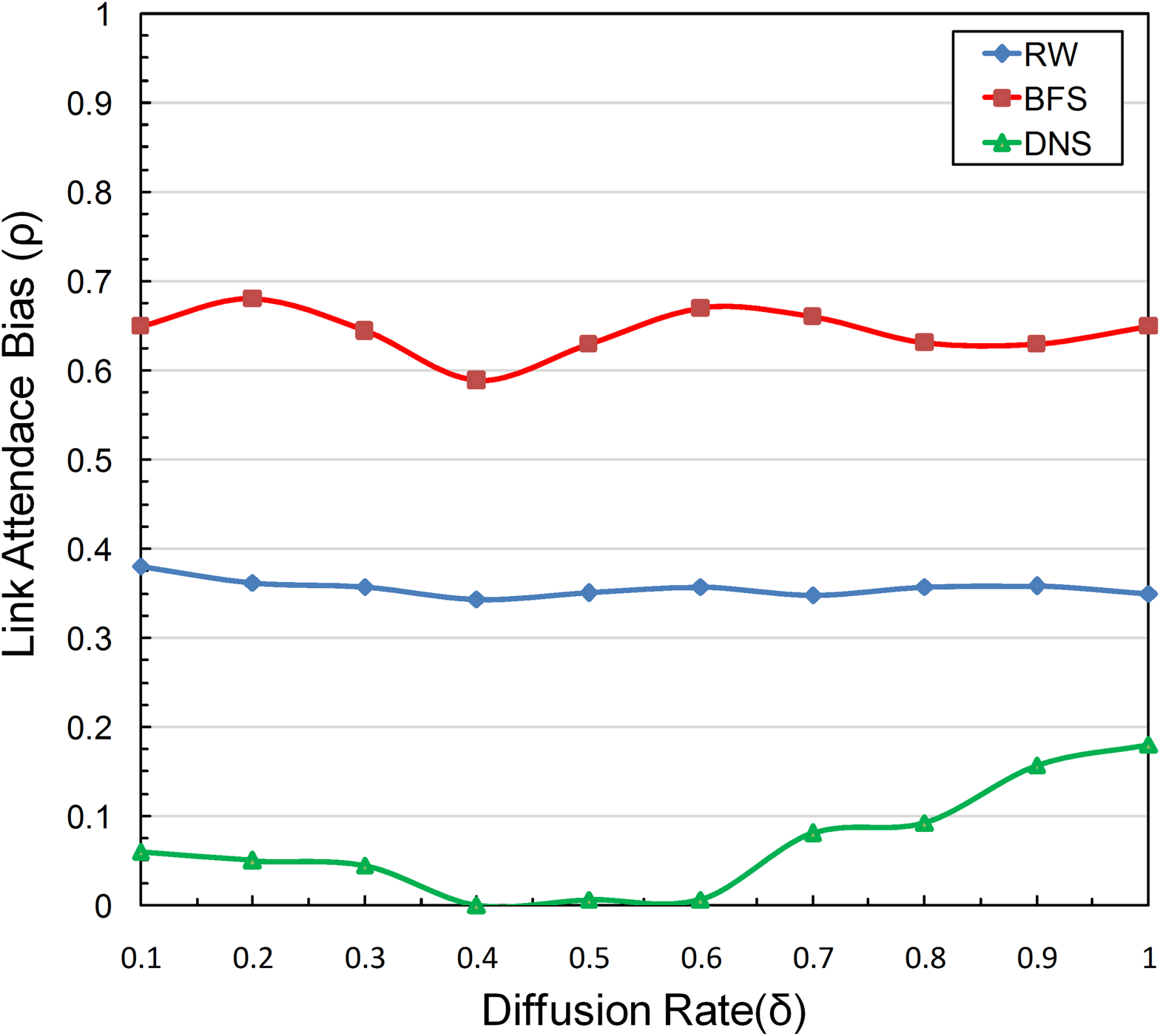}}
    \subfigure[Political Blogsphere Network] {\includegraphics[scale=0.06]{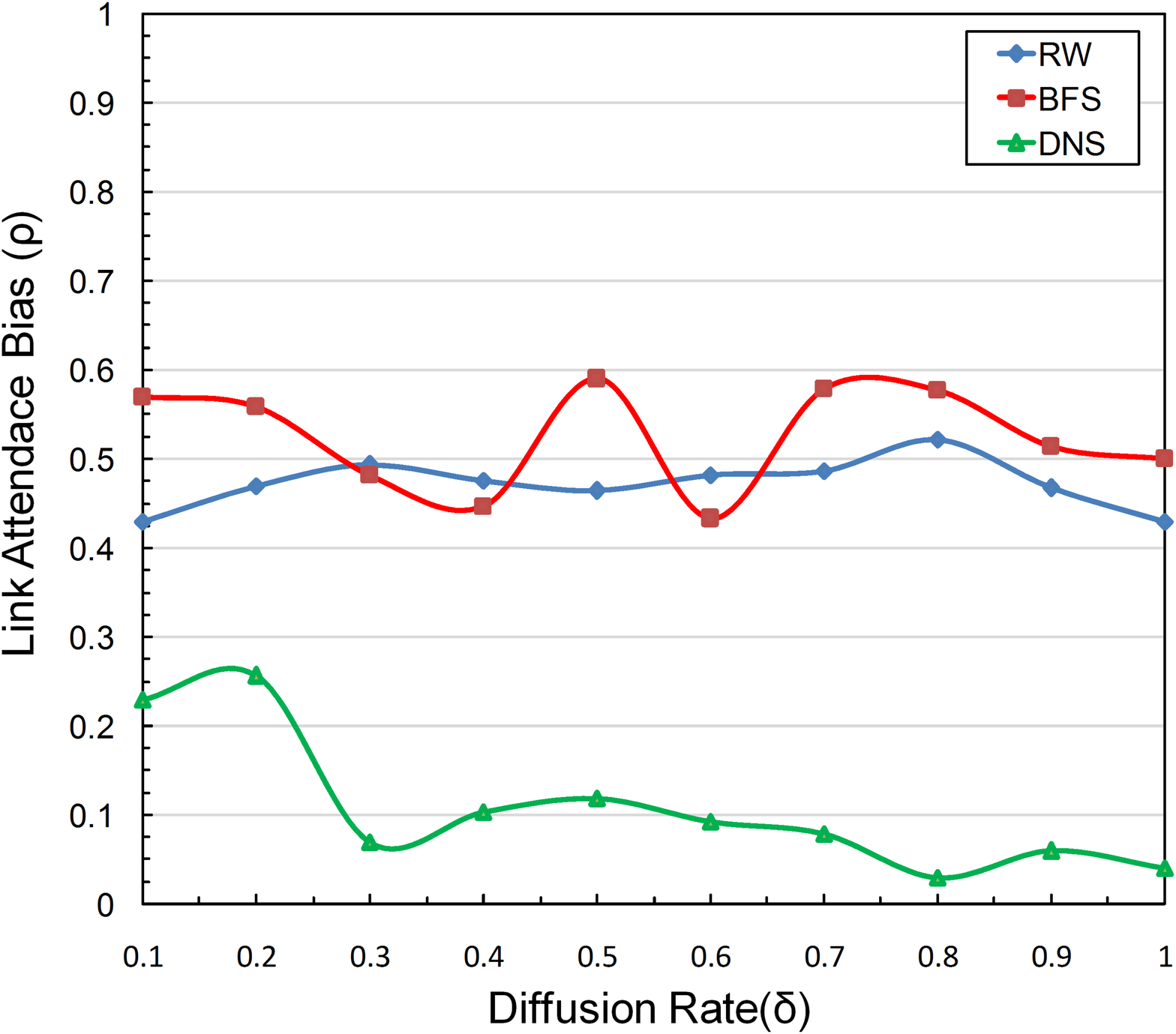}}
    \subfigure[Football Network] {\includegraphics[scale=0.06]{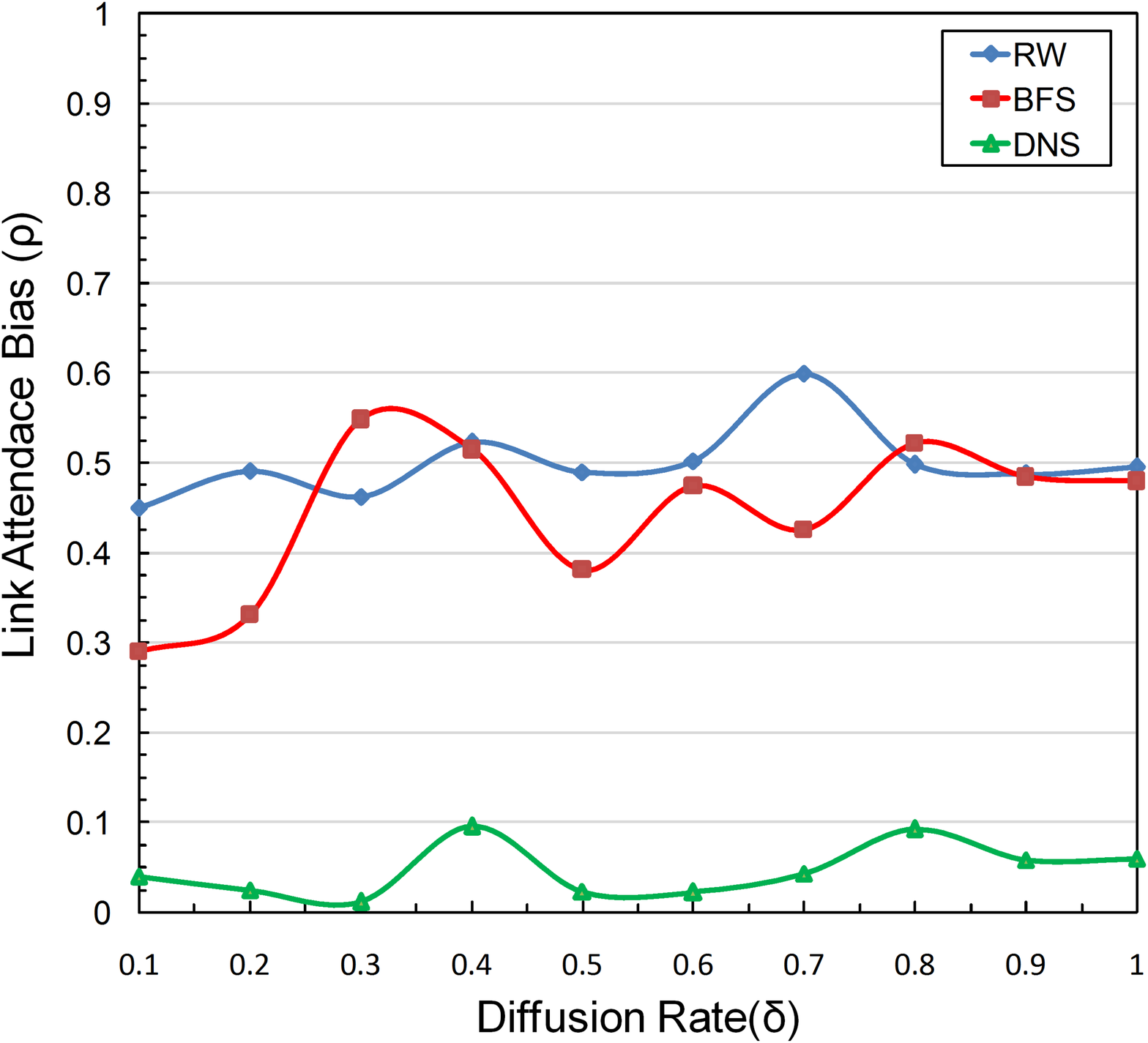}}
     \subfigure[Co-authorship Network]{\includegraphics[scale=0.06]{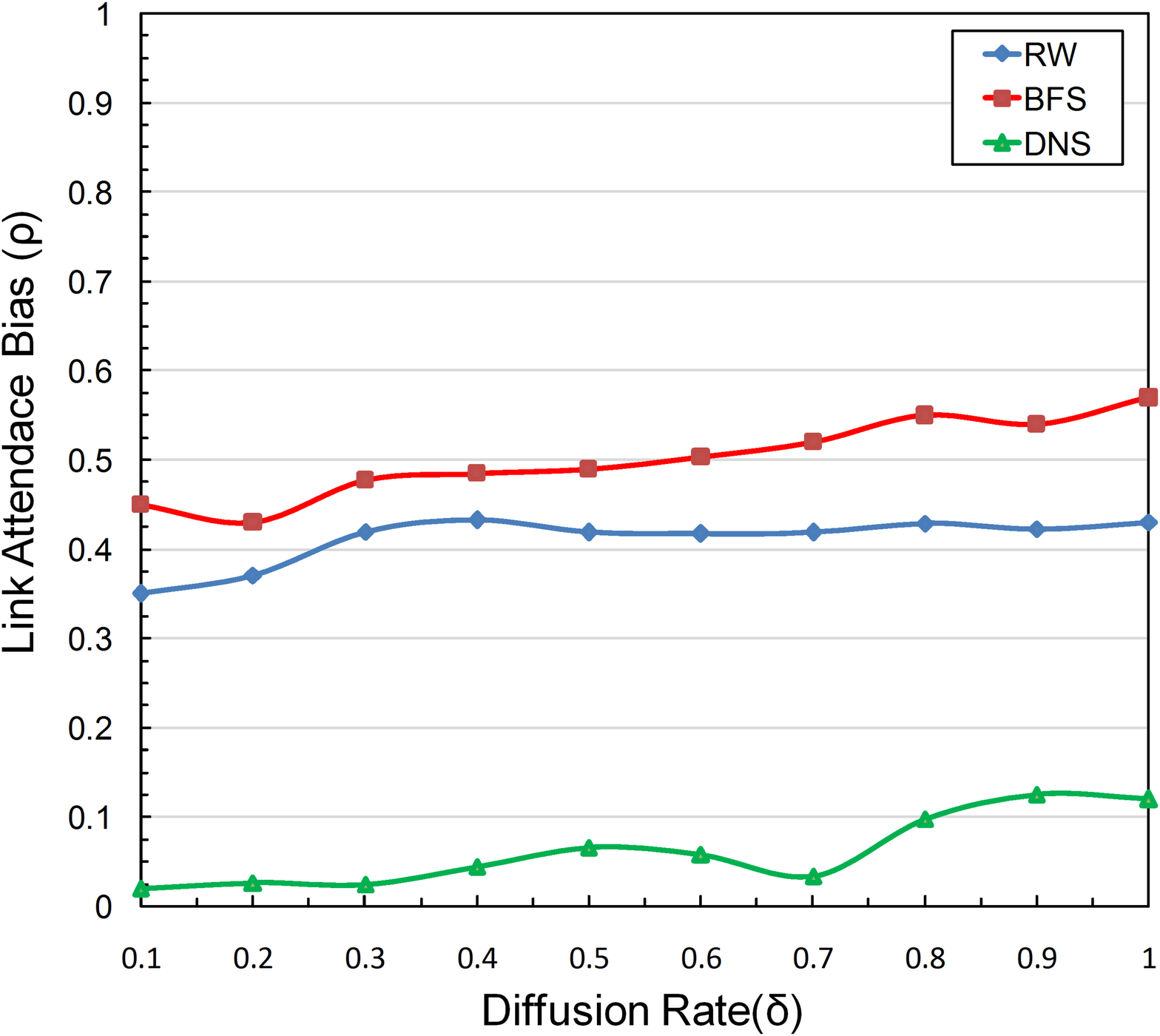}}
  \end{center}
  \caption{Analysis of diffusion rate over sampling frameworks}
  \label{DiffRate}
\end{figure*}

The diffusion rate ($\delta$) of infection over the underlying network has a significant role in gathering diffusion data. As this rate decreases, the smaller parts of the underlying network will be covered by the infection. Therefore, collecting the diffusion data becomes more difficult. Here, we analyze the effect of diffusion rate against performance of the proposed method. Figure \ref{DiffRate} illustrates that \textsc{Dns} leads to low bias even in low diffusion rates. Additionally, these results demonstrates the independence of the proposed framework to the diffusion process behavior. Hence, we can use \textsc{Dns} in various diffusion networks with different diffusion patterns without any loss in performance.

\section{Conclusions}\label{Conclusions}
In this paper, we introduced a novel two-step framework, \textsc{Dns}, to measure the characteristics of large scale and latent diffusion networks. We proposed a sampling algorithm that samples the underlying network by moving from a node to one of its neighboring nodes through an outgoing link by considering the infection probability. Moreover, we proposed three estimators for correcting the bias of sampled data by extending the well-known Hansen-Hurwitz estimator. To this end, we computed the visiting probabilities of three types of diffusion characteristics; link-based, node-based, and cascade-based.

Our experiments showed that in average, the proposed method outperforms BFS and RW in terms of link-attendance by about $37\%$ and $35\%$, respectively. Moreover, we showed that the proposed estimator can improve the performance of the sampling design by about $30\%$. Therefore, an appropriate estimator plays an important role in correcting the bias. Furthermore, the results demonstrated that the proposed method can act very well even in low sampling rates. Additionally, our studies on the diffusion process behavior showed that \textsc{Dns} leads to low bias even in low diffusion rates. 

we believe that our results provide a promising step towards understanding the sampling approaches in analysis and evaluation of diffusion processes. There are several interesting directions for future work. Approximating the visiting probabilities of node-based and cascade-based characteristics is one of our main future goals.

\section{Acknowledgments}
This research has been partially supported by ITRC (Iran Telecommunication Research Center) under grant number 6479/500 (90/4/22).

\end{document}